\newif\ifdraft\drafttrue{}
\newif\iflater\latertrue{}
\newif\ifdesperateforspace\desperateforspacefalse{}
\newif\ifaftersubmission\aftersubmissionfalse{}

\PassOptionsToPackage{names,dvipsnames}{xcolor}
\documentclass[acmsmall,screen,nonacm]{acmart}
\usepackage[none]{hyphenat}

\colorlet{cb-blue}{RoyalBlue}
\colorlet{cb-green}{ForestGreen}
\colorlet{cb-pink}{CarnationPink}
\colorlet{cb-red}{RedOrange}

\usepackage[frozencache,cachedir=minted-cache]{minted}
\setminted{fontsize=\footnotesize, breaklines}

\usepackage{acronym}
\usepackage{changepage}
\usepackage{wrapfig}
\usepackage[justification=centering]{caption}
\usepackage{subcaption}
\usepackage{enumitem}
\usepackage{dblfloatfix} 
\usepackage{graphicx}

\graphicspath{ {./images/} }

\definecolor{dkred}{rgb}{0.7,0,0}
\definecolor{dkblue}{rgb}{0,0,0.7}
\definecolor{notered}{rgb}{0.85,0,0}
\definecolor{dkpurple}{HTML}{4e02eb}
\definecolor{dkgreen}{HTML}{006329}
\definecolor{teal}{HTML}{007982}
\definecolor{fuchsia}{HTML}{8C368C}

\newcommand{\mica}{\textsc{Mica}}
\newcommand{\tyche}{\textsc{Tyche}}

\theoremstyle{definition}

\AtBeginDocument{%
  \providecommand\BibTeX{{%
    \normalfont B\kern-0.5em{\scshape i\kern-0.25em b}\kern-0.8em\TeX}}}

\setcopyright{none}

\title{\mica{}: Automated Differential Testing for OCaml Modules}

\author{Ernest Ng}
\email{eyn5@cornell.edu}
\orcid{0009-0007-3706-6336}
\affiliation{
  \institution{University of Pennsylvania and Cornell University}
  \country{USA}
}
\authornote{Work done while at the University of Pennsylvania.}

\author{Harrison Goldstein}
\email{hgo@seas.upenn.edu}
\orcid{0000-0001-9631-1169}
\affiliation{
  \institution{University of Pennsylvania and University of Maryland}
  \country{USA}
}
\authornotemark[1]

\author{Benjamin C. Pierce}
\email{bcpierce@cis.upenn.edu}
\orcid{0000-0001-7839-1636}
\affiliation{
  \institution{University of Pennsylvania}
  \country{USA}
}

\authorsaddresses{}

\begin{document}
\maketitle
\vspace{-10pt}
\section{Introduction}\label{sec:intro}
Suppose we are given two OCaml modules implementing the same signature. How do we check that they are \textit{observationally equivalent}---that is, that they behave the same on all inputs?  
One established way is to use a \textit{property-based testing} (PBT) tool such as QuickCheck~\cite{claessen_quickcheck_2000}. 
Currently, however, this can require significant amounts of boilerplate code and ad-hoc test harnesses~\cite{goldstein_et_al_icse24}.

\begin{figure}[!b]
\centering
\begin{minipage}[!htp]{0.4\linewidth}
\centering
\begin{minted}[escapeinside=<>, mathescape=true]{ocaml}
(* User code *)

(* Signature for finite sets *)
module type S = sig
  type 'a t
  val empty  : 'a t
  val insert : 'a <$\to$> 'a t <$\to$> 'a t
  ...
end
[@@deriving mica, ...]

(* Modules under test *)
module ListSet : S = ...
module BSTSet  : S = ...

(* Users invoke Mica's test harness on 
   the modules they wish to test *)
module T = TestHarness(ListSet)(BSTSet) 
let () = T.run_tests ()
\end{minted}
\end{minipage}
\hfill\vline\hfill
\begin{minipage}[!htp]{0.55\linewidth}
\centering
\begin{minted}[escapeinside=<>, mathescape=true]{ocaml}
(* Code produced by Mica *)
(* Symbolic expressions *)
type expr = Empty | Insert of int * expr | ...
type ty = Int | IntT | ...

(* QuickCheck generator for [expr]s *)
let rec gen_expr : ty <$\to$> expr Generator.t = ...

(* Interpretation functor *)
module Interpret (M : S) = struct 
  type value = ValInt of int | ValIntT of int M.t | ...

  (* Interprets an [expr] over module [M] *)
  let rec interp : expr <$\to$> value = ...
end

(* Functor for differential testing of [M1] & [M2] *)
module TestHarness (M1 : S) (M2 : S) = struct 
   let run_tests : unit <$\to$> unit = ...
\end{minted}
\end{minipage}
\caption{Left: User code (note the annotation on signature \texttt{S}). Right: PBT code automatically derived by \mica{}.}
\label{fig:intro}
\end{figure}

We present \mica{}, an automated tool for testing equivalence of OCaml modules. \mica{} is implemented as a PPX compiler extension \cite{rebours_ppx_ocaml19}, allowing users to supply minimal annotations to a module signature. These annotations guide \mica{} to automatically derive specialized PBT code that checks observational equivalence using Jane Street's \texttt{Core.Quickcheck} library \cite{eastlund_core_quickcheck}. A \mica{} prototype 
is available on GitHub;\footnote{\url{https://github.com/ngernest/mica}} we are currently reimplementing \mica's concrete syntax as a PPX extension (as described below).

\section{Design of \mica{}}
Suppose we have two modules \texttt{ListSet} and \texttt{BSTSet} that implement finite sets (signature \texttt{S}) using lists and binary search trees (BSTs), respectively. To test for observational equivalence, users invoke \mica{} by annotating \texttt{S} with the directive \texttt{[@@deriving mica]}. During compilation, \mica{} derives the definition for an inductively-defined algebraic data type (ADT) called \texttt{expr}, which represents \textit{symbolic expressions}. Each declaration in \texttt{S} corresponds to a constructor for the \texttt{expr} ADT with the same name, arity and argument types. \mica{} also derives auxiliary ADTs that represent the possible \textit{types} and \textit{values} of symbolic expressions. 

To generate random symbolic expressions, \mica{} derives a recursive QuickCheck generator \texttt{gen\_expr} that is parameterized by the desired type of the expression. The type-directed nature of this generator ensures that only {well-typed} expressions are produced. Subsequently, to interpret symbolic expressions over a specific module \texttt{M} and produce concrete \texttt{value}s, \mica{} produces an interpretation functor that is parameterized by an instance of \texttt{S}.  

To check for observational equivalence, \mica{} produces a functor \texttt{TestHarness} which users instantiate with the desired modules. Crucially, \mica{}'s test harness only compares the \texttt{value} of interpreted \texttt{expr}s at \textit{concrete types}, for example \texttt{int}, not the abstract type \texttt{'a} \texttt{t}, since the internal representations of such values may differ arbitrarily. 

To test modules with mutable internal state, the \texttt{expr} datatype is extended with a constructor \texttt{Seq}, where \texttt{Seq(e1,} \texttt{e2)} represents the \textit{sequencing} of expressions \texttt{e1} and \texttt{e2}. 
Also, we are currently working on extending \mica{} with the ability to derive constructors that represent \texttt{let}-expressions. This addition will allow \texttt{expr}s to refer to previously generated data, encoding dependencies between successive function calls.

To test polymorphic functions, \mica{} instantiates all type variables \texttt{'a} with \texttt{int}, following well-known heuristics \cite{bernady_et_al_2010, favonia_wang_2022}. Additionally, \mica{} offers support for generating unary anonymous functions. For example, to test the polymorphic higher-order function \texttt{map}, \mica{} derives the symbolic expression \texttt{Map} \texttt{of} \texttt{(int} $\to$ \texttt{int)} \texttt{*} \texttt{expr}, generating a random \texttt{int} $\to$ \texttt{int} function in the process using canonical techniques from the PBT literature \cite{claessen_2012, eastlund_core_quickcheck}. 

\begin{wrapfigure}{r}[0pt]{0.36\textwidth}
\vspace{-0.35in}
\includegraphics[width=0.36\textwidth]{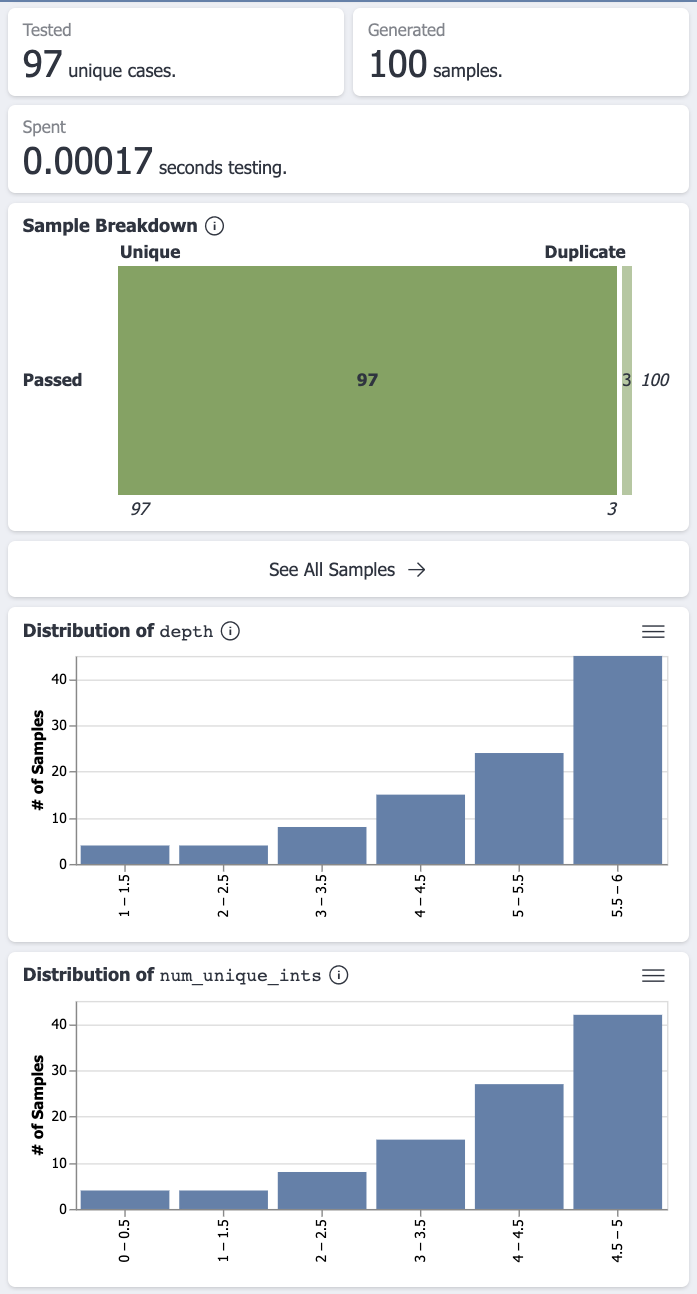}
\caption{The \tyche{} user interface, displaying \mica{}'s test results}
\label{fig:tyche_screenshot}
\vspace{-10mm}
\end{wrapfigure}
\section{Editor Integration}

We have integrated \mica{} with \tyche{} \cite{goldsteinTycheMakingSense2024}, an extension to VSCode for visualizing the behavior of PBT generators. Whenever \mica{} checks two modules for observational equivalence at type $\tau$, \tyche{} plots numeric features regarding the random \texttt{expr}s of type $\tau$ that were used to test the two modules. For example, \tyche{} serializes the individual \texttt{expr}s generated and also visualizes the distribution of their depth, thereby giving users greater insight into the effectiveness of \mica{}'s testing. We refer the reader to \citeauthor{goldsteinTycheMakingSense2024}'s work for further details regarding \tyche{}. 

\vspace{-1mm}
\section{Case Studies}
To examine \mica{}'s efficacy as a testing tool, we applied it with various module signatures that admit multiple implementations, including:
\begin{itemize}
    \item Regular expression matchers (Brzozowski derivatives, deterministic finite automata) \cite{elliott_icfp21, goldstein_livestream}
    \item Jane Street's imperative \texttt{Base.Queue} and \\ \texttt{Base.Linked\_queue} modules \cite{jane_street_base}
    \item Character sets, implemented respectively using the standard library's \texttt{Set.Make(Char)} module and the \texttt{charset} library (a specialized implementation that uses compiler intrinsics for efficiency purposes) \cite{yallop_charsets} 
    \item Polynomials (Horner schema, monomials) \cite{filliatre_polynomials, fletcher_polynomials}
    \item Finite maps (red-black trees, association lists) \cite{okasaki_1999, clarkson_et_al_textbook}
    \item Unsigned 32 $\&$ 64-bit integer arithmetic (the \texttt{stdint} and \texttt{ocaml-integers} libraries) \cite{ocaml_stdint, yallop_integers}
\end{itemize}
\mica{} was able to found 35 manually-inserted bugs inserted across these modules without any user input required. 

We have also replicated a case study from John Hughes's \textit{How to Specify It} \cite{hughes_2020}, an extended tutorial on Haskell QuickCheck which uses BSTs representing finite maps as its running example. (Hughes's paper is a well-known benchmark in the PBT literature \cite{Holey2022, etna_icfp23, prinz_lampropoulos_pldi23, roughspec_ifl2020}.) The paper's accompanying artifact \cite{HowToSpecifyItCode} contains one correct BST implementation and eight erroneous ones. For example, one bug results in a singleton tree being returned during BST insertion, while another bug reverses key comparison when deleting a key-value pair from the tree.\footnote{We refer the reader to Hughes's paper \cite{hughes_2020} for detailed descriptions of all eight bugs.} We ported these implementations to OCaml as nine separate modules. Subsequently, we found that \mica{} was able to successfully detect divergent behavior between the correct and erroneous modules. 

Specifically, we evaluate \mica{} by measuring the average number of tests required to provoke failure in each observational equivalence test. We measure this average by executing the PBT code derived by \mica{} for 1000 times, each time with a different random seed. Following a technique established in prior work \cite{Holey2022}, in all our tests, we generate keys uniformly at random from the range 0 to \texttt{size}, where \texttt{size} is the internal size parameter of \mica{}'s QuickCheck generator. As Figure \ref{fig:time_to_failure_table} shows, \mica{} was able to detect all bugs in Hughes's repository without any user intervention.

\begin{figure}[t]
\begin{tabular}{|c|c|c|c|c|c|c|c|c|}
\hline
\multicolumn{1}{|l|}{\textbf{}} & \textbf{Bug \#1} & \textbf{Bug \#2} & \textbf{Bug \#3} & \textbf{Bug \#4} & \textbf{Bug \#5} & \textbf{Bug \#6} & \textbf{Bug \#7} & \textbf{Bug \#8} \\ \hline
\textbf{Min} & 6 & 8 & 504 & 7 & 42 & 10 & 17 & 20 \\ \hline
\textbf{Mean} & 20 & 62 & 553 & 20 & 286 & 44 & 163 & 229 \\ \hline
\textbf{Max} & 118 & 262 & 765 & 94 & 546 & 238 & 312 & 438 \\ \hline
\end{tabular}
\vspace{-2mm}
\caption{Average mean no. of trials required to provoke failure in an observational equivalence test}
\label{fig:time_to_failure_table}
\vspace{-5mm}
\end{figure}

In addition, we have also used \mica{} to catch bugs in students' homework submissions for the University of Pennsylvania's undergraduate OCaml course \cite{cis1200_website}. As homework, students were asked to implement a signature for finite sets using both ordered lists and BSTs \cite{cis1200_hw3}. Students were also instructed to submit a test suite of unit tests with which they tested their two implementations. After collecting all the students' submissions, we used \mica{} to examine whether their set implementations were observationally equivalent. We observed that \mica{} detected observational equivalence bugs in 29\% of the students' submissions (107 out of 374 students), with most bugs (91\%) caught within 300 randomly-generated inputs. Notably, these bugs were not caught within students' manually-written unit tests, demonstrating \mica{}'s ability to facilitate more robust differential testing.

\section{Related Work}
\textsc{Monolith} \cite{pottier_monolith_2021} and \textsc{Articheck} \cite{braibant_articheck_2014} are differential testing frameworks for ML modules that provide users with GADT-based DSLs to represent well-typed sequences of function calls. Using these DSLs, users declare functions to be tested across modules; both libraries use coverage-guided fuzzers to enumerate inhabitants of abstract data types during testing. Like these tools, \mica{} generates well-typed symbolic expressions, but it obviates the need for users to learn specialized DSLs, automatically producing specialized PBT code instead.

\textit{Model-based testing} is a similar style of testing which examines whether the system under test is observationally equivalent to an abstract model. Model-based testing was pioneered in the PBT community by \textsc{QuviQ}'s Erlang QuickCheck library \cite{quviq_quickcheck_2006}, which uses finite state machines as abstract models. This approach was brought to OCaml via the state-machine based PBT library \textsc{QCSTM} ~\cite{midtgaard_qcstm_2020}. In \textsc{QCSTM}, symbolic expressions are represented as algebraic data types (ADTs), while the testing harness features state-dependent QuickCheck generators for symbolic expressions, along with functions that interpret expressions over both the model and target implementations. Our work builds on \textsc{QCSTM} by utilizing a similar ADT-based representation for symbolic expressions and adding support for testing binary operations over abstract types. 

For testing ML modules more broadly, one can utilize \textsc{Gospel} \cite{chargueraud_gospel_2019}, a specification language for ML modules, along with \textsc{Ortac} \cite{filliatre_ortac_2021}, a runtime assertion-checking tool that checks \textsc{Gospel} specifications. Notably, \textsc{Ortac} offers a plugin that supports \textsc{QCheck-STM} \cite{midtgaard_ocaml_2022}, a variant of \textsc{QCSTM} adapted for testing parallel Multicore OCaml code. In this setup, in addition to generating random symbolic expressions, the test harness also checks whether pre- and post-conditions expressed using \textsc{Gospel} are satisfied in-between function calls \cite{ortac_qcheck_stm_plugin}.

\section{Future Work}
We plan to extend \mica{} to support OCaml functors and modules with multiple abstract types, and add the ability to generate a wider variety of higher-order functions. Furthermore, inspired by recent tools that combine coverage-guided fuzzing and PBT \cite{crowbar_dolan_2017, lampropoulos_oopsla19}, we plan on investigating whether coverage information could be used to tune \mica{}'s generator of random \texttt{expr}s so that newly generated \texttt{expr}s tend to exercise previously untested code. 

Finally, although the PBT code derived by \mica{} currently uses Jane Street's \texttt{Core.Quickcheck} library, \mica{}'s design is library-agnostic. We leave it as future work to adapt \mica{} to support other OCaml PBT frameworks (e.g. QCheck \cite{cruanes_qcheck_2013}), building on recent work that uses the \textsc{Etna} PBT evaluation platform \cite{etna_icfp23} to compare the efficacy of different OCaml PBT frameworks \cite{kamath_pldi24src}.

\section*{Acknowledgements}
A prototype of \mica{} was presented at the ICFP '23 Student Research Competition; we thank the ICFP SRC program committee for their feedback. We also thank Jeremy Yallop, Carl Eastlund, David Vulakh, and Jan Midtgaard for comments and suggestions. Lastly, we thank the University of Pennsylvania's PLClub for support and encouragement. Development of \mica{} was supported by the National Science Foundation under grant NSF \#2402449, {\em SHF: Medium: Usable Property-Based Testing}.

\bibliographystyle{ACM-Reference-Format}
\bibliography{references}
\end{document}